\documentclass[a4paper,11pt]{article}
\pdfoutput=1 

\usepackage{jheppub} 

\usepackage[T1]{fontenc} 

\title{\boldmath Wave Equation for the Wu Black Hole}

\author[a,b]{Tolga Birkandan}
\author[a,b,c]{Mirjam Cveti\v c,}

\affiliation[a]{Department of Physics and Astronomy, University of Pennsylvania, \\Philadelphia, PA 19104, USA.}
\affiliation[b]{Department of Physics, Istanbul Technical University, \\Istanbul 34469, Turkey.}
\affiliation[c]{Center for Applied Mathematics and Theoretical Physics, University of Maribor, \\Maribor, Slovenia.}

\emailAdd{birkandant@itu.edu.tr}
\emailAdd{cvetic@hep.upenn.edu}

\abstract{
Wu black hole is the most general solution of maximally supersymmetric gauged supergravity in D=5, containing $U(1)^{3}$ gauge symmetry. We study the separability of the  massless Klein-Gordon equation and probe its singularities for a general stationary, axisymmetric metric with orthogonal transitivity, and apply the results to the Wu black hole solution. We start with the zero azimuthal-angle eigenvalues in the scalar field Ansatz and find that the residuum of a pole in the radial equation is associated with the surface gravity calculated at this horizon. We then generalize our calculations to nonzero azimuthal eigenvalues and probing each horizon singularity, we show that the residua of the singularities for each horizon are in general associated  with  a specific combination of the surface gravity and the angular velocities at the associated horizon. It turns out that for the Wu black hole both the radial and angular equations are general Heun's equations  with four regular singularities.
}

\begin{document}
\preprint{UPR-1263-T}
\maketitle
\flushbottom
\section{Introduction}
The microscopic treatment of black holes shows that all horizons of the black holes contribute in calculations, unlike the macroscopic classical theory which regards only the outermost horizon, namely the event horizon. As a result of this fact, scattering calculations should involve probing the vicinity of the horizons and the structure one obtains is the key element of the scattering phenomena and gauge/gravity correspondence.

The literature with the separability analysis of the wave equations and the subtracted geometry of the asymptotically flat solutions \cite{Strominger:1996sh,CYII,CYI,CYIII,L,CLI,CLII,CLIII,CR,CMS,GHSS,CL11I,CL11II,CG,liulu} is much richer and complete than the number of studies on the asymptotically anti-de Sitter (AdS) spacetimes, including the recent most general solution for the maximally supersymmetric ungauged supergravity \cite{Chow:2013tia,Chow:2014cca,Cvetic:2014sxa}. The remarkable structure of separability seen in these works has also been found in some gauged supergravity solutions \cite{cclp,Chong05,Chong:2006zx,Chong:2005da,Chong:2004dy,Birkandan:2011fr,Aliev:2008yk,Wu:2009ug}. The symmetries of various black hole solutions in supergravity was studied in \cite{Chow:2008fe}.

Deriving the Klein-Gordon equation, studying its separability and probing the residua associated with the horizons are the key issues that constitutes the core of the calculations for the black hole internal structure. One finds the physical properties, namely surface gravities and angular velocities associated with the horizons by probing the radial part of the Klein-Gordon equation. As the surface gravity carry the information of the Hawking temperature on the horizon, we see the thermodynamics explicitly in the wave equation for each horizon. The techniques used for finding the exponents which form the monodromy matrices seem to be helpful also in probing the radial wave equation to obtain the residua at each pole of the horizon equation \cite{Castro:2013lba,Castro:2013kea,Birkandan:2014cva}.

Based on the studies of asymptotically flat solutions \cite{L,CLI,CLII,CR}, recent studies show  the universal nature of the area products and entropy products at all horizons of black holes in gauged supergravities and higher derivative gravity theories,  as  they depend only on the quantized charges, quantized angular momenta and cosmological constant \cite{Cvetic:2010mn, Castro:2013pqa,Cvetic:2013eda}. This strong condition also enhances the importance of probing the horizons carefully.

The  goal of this paper is to study the separability of the massless scalar wave equation for a general stationary, axisymmetric metric with orthogonal transitivity and apply the results to the most general black hole solution in D=5 maximally supersymmetric gauged supergravity found by Wu \cite{Wu:2011gq}. While we derive the  residua structure for a general stationary, axisymmetric metric with orthogonal transitivity,  due to its complexity in D=5 with two angular momenta, three charges, and cosmological constant, Wu's solution is an excellent testing ground to study explicitly  properties and structure of the wave equation; in particular the separability and the structure of the poles.  Probing each horizon singularity, we  show that the residua of the singularities for each horizon are associated in  general form with the surface gravity and the angular velocities at the associated horizon. Our calculations  show that the separability crucially rely on the determinant of the metric is of a  simple form and regular at the horizon.

We will briefly introduce the Wu's solution of the general nonextremal rotating charged AdS black hole in five-dimensional $U(1)^{3}$ gauged supergravity in the next section. The section three covers the Klein-Gordon equation, its separation and an analysis on the radial part which brings it in a form that uncovers the thermodynamics of each horizon explicitly with associated surface gravities for zero eigenvalues for the azimuthal coordinates in the scalar field Ansatz. The section four generalizes our calculation to the non-zero azimuthal eigenvalues and the appendices include the key parts of the separation process in detail.
\section{The metric}
In \cite{Wu:2011gq}, Wu's original solution of the general nonextremal rotating charged AdS black hole in five-dimensional $U(1)^{3}$ gauged supergravity is given as
\begin{eqnarray}
ds^{2} &=&(H_{1}H_{2}H_{3})^{1/3}\left[ -\frac{(1+g^{2}r^{2})\Delta _{\theta
}}{\xi _{a}\xi _{b}}\,dt^{2}+\Sigma \left( \frac{r^{2}\,dr^{2}}{\Delta _{r}}+%
\frac{d\theta ^{2}}{\Delta _{\theta }}\right) \right. +\frac{%
(r^{2}+a^{2})\sin ^{2}\theta }{\xi _{a}}\,d\phi ^{2} \notag\\
&&+\frac{(r^{2}+b^{2})\cos^{2}\theta }{\xi _{b}}\,d\psi ^{2}
 +\frac{2ms_{1}^{2}}{\Sigma
H_{1}(s_{1}^{2}-s_{2}^{2})(s_{1}^{2}-s_{3}^{2})}K_{1}^{2} \notag\\
&&+\frac{2ms_{2}^{2}}{%
\Sigma H_{2}(s_{2}^{2}-s_{1}^{2})(s_{2}^{2}-s_{3}^{2})}K_{2}^{2}\left. +%
\frac{2ms_{3}^{2}}{\Sigma H_{3}(s_{3}^{2}-s_{1}^{2})(s_{3}^{2}-s_{2}^{2})}%
K_{3}^{2}\right] ,
\end{eqnarray}
where
\begin{eqnarray}
K_{i} &=&\frac{s_{i}c_{1}c_{2}c_{3}}{c_{i}}\frac{\sqrt{\Xi _{1a}\Xi _{2a}\Xi
_{3a}\Xi _{1b}\Xi _{2b}\Xi _{3b}}}{\sqrt{\Xi _{ia}\Xi _{ib}}} \notag\\
&&\times \left( \frac{c_{i}^{2}}{c_{1}c_{2}c_{3}}\frac{\Delta _{\theta }}{\xi _{a}\xi _{b}}\,dt-%
\frac{\Xi _{ia}}{\sqrt{\Xi _{1a}\Xi _{2a}\Xi _{3a}}}\frac{a\sin ^{2}\theta }{%
\xi _{a}}\,d\phi -\frac{\Xi _{ib}}{\sqrt{\Xi _{1b}\Xi _{2b}\Xi _{3b}}}\frac{%
b\cos ^{2}\theta }{\xi _{b}}\,d\psi \right)  \notag \\
&&+\frac{c_{i}s_{1}s_{2}s_{3}}{s_{i}}\sqrt{\Xi _{ia}\Xi _{ib}} \\
&&\times \left( -\frac{%
c_{1}c_{2}c_{3}}{c_{i}^{2}}\frac{g^{2}ab\Delta _{\theta }}{\xi _{a}\xi _{b}}%
\,dt+\frac{\sqrt{\Xi _{1a}\Xi _{2a}\Xi _{3a}}}{\Xi _{ia}}\frac{b\sin
^{2}\theta }{\xi _{a}}\,d\phi +\frac{\sqrt{\Xi _{1b}\Xi _{2b}\Xi _{3b}}}{\Xi
_{ib}}\frac{a\cos ^{2}\theta }{\xi _{b}}\,d\psi \right) .\notag
\end{eqnarray}
or
\begin{equation}
K_{i}=K_{it}(\theta)dt+K_{i\phi }(\theta)d\phi +K_{i\psi }(\theta)d\psi ,
\end{equation}
in a compact form showing the coordinate dependencies. The parameters are
\begin{eqnarray}
\Xi _{ia} &=&c_{i}^{2}-s_{i}^{2}\xi _{a},\text{ \ \ }\Xi
_{ib}=c_{i}^{2}-s_{i}^{2}\xi _{b}, \\
\xi _{a} &=&1-g^{2}a^{2},\text{ \ \ \ }\xi _{b}=1-g^{2}b^{2}, \\
s_{i} &=&\sinh \delta _{i},\text{ \ \ \ \ }c_{i}=\cosh \delta _{i}, \\
\Sigma &=&r^{2}+a^{2}\cos ^{2}\theta +b^{2}\sin ^{2}\theta , \\
\Delta _{\theta } &=&1-g^{2}\left( a^{2}\cos ^{2}\theta +b^{2}\sin
^{2}\theta \right) , \\
H_{i} &=&1+\frac{2ms_{i}^{2}}{\Sigma },
\end{eqnarray}
and the horizon equation is written as
\begin{eqnarray}
\Delta _{r}
&=&(r^{2}+a^{2})(r^{2}+b^{2})(1+g^{2}r^{2})-2mr^{2}+2mg^{2}\left\{
(s_{1}^{2}+s_{2}^{2}+s_{3}^{2})r^{4}\right.  \notag \\
&&-(s_{1}^{2}s_{2}^{2}+s_{1}^{2}s_{3}^{2}+s_{2}^{2}s_{3}^{2})\left[
(a^{2}+b^{2}-2m)r^{2}+a^{2}b^{2}(2+g^{2}r^{2})\right]  \notag \\
&&+s_{1}^{2}s_{2}^{2}s_{3}^{2}\left(
[(a+b)^{2}-2m][(a-b)^{2}-2m]-2g^{2}a^{2}b^{2}(2r^{2}+2m\right.  \notag \\
&&\left. +a^{2}+b^{2})+g^{4}a^{4}b^{4}\right)
+2mg^{2}a^{2}b^{2}[s_{1}^{4}s_{2}^{4}+s_{1}^{4}s_{3}^{4}+s_{2}^{4}s_{3}^{4}
\notag \\
&&\left.
-2s_{1}^{2}s_{2}^{2}s_{3}^{2}(s_{1}^{2}+s_{2}^{2}+s_{3}^{2})]\right\}.
\end{eqnarray}
Let us use the transformation
\begin{eqnarray}
u &\equiv &r^{2}, \\
y &\equiv &a^{2}\cos ^{2}\theta +b^{2}\sin ^{2}\theta ,
\end{eqnarray}
to see the singularity structure better. Then the metric becomes
\begin{eqnarray}
ds^{2} &=&(H_{1}H_{2}H_{3})^{1/3}\left[ -\frac{(1+g^{2}u)Y}{\xi _{a}\xi
_{b}(a^{2}-y)(y-b^{2})}\,dt^{2}+\Sigma \left( \frac{\,du^{2}}{4X}+\frac{%
dy^{2}}{4Y}\right) \right. \notag\\
&&+\frac{(u+a^{2})(a^{2}-y)}{\xi _{a}(a^{2}-b^{2})}%
\,d\phi ^{2}+\frac{(u+b^{2})(y-b^{2})}{\xi _{b}(a^{2}-b^{2})}\,d\psi ^{2}
\notag \\
&&\quad +\frac{2m\eta _{1}}{\Sigma H_{1}}K_{1}^{2}+\frac{2ms\eta _{2}}{%
\Sigma H_{2}}K_{2}^{2}\left. +\frac{2m\eta _{3}}{\Sigma H_{3}}K_{3}^{2}%
\right].
\end{eqnarray}
Here, the parameters are defined as
\begin{eqnarray}
\Sigma &=&u+y, \\
X &=&\Delta _{r \to u}=g^{2}(u-u_{1})(u-u_{2})(u-u_{3}), \label{eq:horeqn}\\
Y &=&(a^{2}-y)(y-b^{2}) \Delta _{\theta \to y}=(a^{2}-y)(y-b^{2})(1-g^{2}y),
\end{eqnarray}
and the coefficients of the $K_i$ terms are shortened using
\begin{equation}
\eta _{1}=\frac{s_{1}^{2}}{(s_{1}^{2}-s_{2}^{2})(s_{1}^{2}-s_{3}^{2})},\text{
}\eta _{2}=\frac{s_{2}^{2}}{(s_{2}^{2}-s_{1}^{2})(s_{2}^{2}-s_{3}^{2})},%
\text{ }\eta _{3}=\frac{s_{3}^{2}}{(s_{3}^{2}-s_{1}^{2})(s_{3}^{2}-s_{2}^{2})%
}.
\end{equation}
This transformation with the new parameters will be used in the separation process which will be discussed in the following sections.

The entropy and the surface gravity, following \cite{Wu:2011gq}, at a horizon $u=u_i$ take the form
\begin{equation}
S_i=\frac{\pi ^{2}}{2\xi _{a}\xi _{b}}\sqrt{\frac{W_{i}}{u_{i}}}, \label{eq:entropy}
\end{equation}
and
\begin{equation}
\kappa _{i}=2\pi T_{i}=\sqrt{\frac{u_{i}}{W_{i}}}\left( \frac{dX}{du}\right)|_{u=u_{i}}, \label{eq:kappas}
\end{equation}
where
\begin{eqnarray}
W_{i}&=&[(u_{i}+a^{2})(u_{i}+b^{2})+2mu_{i}(s_{1}^{2}+s_{2}^{2}+s_{3}^{2})]\{(u_{i}+a^{2})(u_{i}+b^{2})  \notag \\
&&+2mg^{2}[(a+b)^{2}-g^{2}a^{2}b^{2}][(a-b)^{2}-g^{2}a^{2}b^{2}]s_{1}^{2}s_{2}^{2}s_{3}^{2}-4mg^{2}a^{2}b^{2}(s_{1}^{2}s_{2}^{2}\notag \\
&&+s_{1}^{2}s_{3}^{2}+s_{2}^{2}s_{3}^{2})\}+8m^{2}u_{i}c_{1}c_{2}c_{3}s_{1}s_{2}s_{3}ab\sqrt{\Xi _{1a}\Xi _{2a}\Xi_{3a}\Xi _{1b}\Xi _{2b}\Xi _{3b}}  \notag \\
&&+4m^{2}u_{i}[u_{i}+g^{2}a^{2}b^{2}(s_{1}^{2}+s_{2}^{2}+s_{3}^{2})](s_{1}^{2}s_{2}^{2}+s_{1}^{2}s_{3}^{2}+s_{2}^{2}s_{3}^{2})\notag \\
&&-4m^{2}\{(a^{2}+b^{2})(1+g^{2}a^{2})(1+g^{2}b^{2})u_{i}+g^{2}[(a^{4}+b^{4})u_{i}\notag \\
&&+g^{2}a^{4}b^{4}(2+g^{2}u_{i})](s_{1}^{2}+s_{2}^{2}+s_{3}^{2})+g^{2}a^{2}b^{2}(a^{2}+b^{2}\notag \\
&&+g^{2}a^{2}b^{2})[2+g^{2}u_{i}(s_{1}^{2}s_{2}^{2}+s_{1}^{2}s_{3}^{2}+s_{2}^{2}s_{3}^{2})]\}s_{1}^{2}s_{2}^{2}s_{3}^{2}\notag \\
&&+4m^{2}g^{4}a^{4}b^{4}(s_{1}^{4}s_{2}^{4}+s_{1}^{4}s_{3}^{4}+s_{2}^{4}s_{3}^{4})-8m^{2}u_{i}g^{6}a^{4}b^{4}s_{1}^{4}s_{2}^{4}s_{3}^{4}\notag \\
&&+8m^{3}(u_{i}+g^{2}a^{2}b^{2})s_{1}^{2}s_{2}^{2}s_{3}^{2}\,,
\end{eqnarray}
and the azimuthal angular velocities are
\begin{eqnarray}
\Omega_{\phi i} &=& \frac{2mu_i}{W_i}\bigg\{ac_1c_2c_3\sqrt{\Xi_{1a}\Xi_{2a}\Xi_{3a}}\Big[u_i+b^2
 -2mg^2b^2\big(s_1^2s_2^2+s_2^2s_3^2 \nonumber \\
&& +s_1^2s_3^2 +2g^2b^2s_1^2s_2^2s_3^2\big)\Big]
 -bs_1s_2s_3\sqrt{\Xi_{1b}\Xi_{2b}\Xi_{3b}}\Big[(u_i+b^2)\xi_a^2 \nonumber \\
&& -2m\big[1+g^2a^2 +2g^2a^2(s_1^2+s_2^2+s_3^2) +g^2a^2(s_1^2s_2^2+s_2^2s_3^2 \nonumber \\
&& +s_1^2s_3^2)(1+g^2a^2) +2g^4a^4s_1^2s_2^2s_3^2\big]\Big]\bigg\} \, , \label{eq:Omegaa}\\
\Omega_{\psi i} &=& \frac{2mu_i}{W_i}\bigg\{bc_1c_2c_3\sqrt{\Xi_{1b}\Xi_{2b}\Xi_{3b}}\Big[u_i+a^2
 -2mg^2a^2\big(s_1^2s_2^2+s_2^2s_3^2 \nonumber \\
&& +s_1^2s_3^2 +2g^2a^2s_1^2s_2^2s_3^2\big)\Big]
 -as_1s_2s_3\sqrt{\Xi_{1a}\Xi_{2a}\Xi_{3a}} \Big[(u_i+a^2)\xi_b^2 \nonumber \\
&& -2m\big[1+g^2b^2 +2g^2b^2(s_1^2+s_2^2+s_3^2) +g^2b^2(s_1^2s_2^2+s_2^2s_3^2 \nonumber \\
&& +s_1^2s_3^2)(1+g^2b^2) +2g^4b^4s_1^2s_2^2s_3^2\big]\Big]\bigg\} \label{eq:Omegab}\, ,
\end{eqnarray}
at the horizon $u=u_i$.
\section{Separation of the Klein-Gordon equation}
The massless Klein-Gordon equation
\begin{equation}
\frac{1}{\sqrt{-g}}\partial _{\mu}\left(\sqrt{-g}g^{\mu \nu }\partial _{\nu }\Phi \right)=0, \label{eq:kg}
\end{equation}
can be written with the scalar field Ansatz
\begin{equation}
\Phi =e^{-i\omega t}e^{im_{1}\phi }e^{im_{2}\psi }R(u)S(y). \label{eq:ansatz}
\end{equation}
We start with this Ansatz to show the separability of the massless Klein-Gordon equation and the validity of the Ansatz will be justified post-factum. The separability of the Klein-Gordon equation for a stationary and axisymmetric space with orthogonal transitivity is guaranteed by the separation of the entire inverse metric elements multiplied by a conformal factor $\Omega_{con}$ \cite{Keeler:2012mq}. This factor turns out to be related to $\sqrt{-g}$ in our case as
\begin{equation}
\Omega_{con}=(H_{1}H_{2}H_{3})^{1/3}\Sigma=4\xi _{a}\xi_{b}(a^{2}-b^{2})\sqrt{-g}.
\end{equation}
The details of the separation process can be found in Appendix A. If we take zero azimuthal eigenvalues $m_{1}=m_{2}=0$, the form of the Klein-Gordon equation is
\begin{equation}
\frac{1}{\sqrt{-g}}\left[ \partial _{t}\left( \sqrt{-g}g^{tt}\partial
_{t}\Phi \right) +\partial _{u}\left( \sqrt{-g}g^{uu}\partial _{u}\Phi
\right) +\partial _{y}\left( \sqrt{-g}g^{yy}\partial _{y}\Phi \right) %
\right] =0.
\end{equation}
The calculation will be generalized to nonzero azimuthal eigenvalues $m_{1} \neq m_{2} \neq 0$ in the next section. Let us write this expression more explicitly as
\begin{eqnarray}
-\omega ^{2}\sqrt{-g}g^{tt}+\left( \sqrt{-g}g^{uu}\right) \frac{(\partial
_{uu}R)}{R}+\left[ \partial _{u}\left( \sqrt{-g}g^{uu}\right) \right] \frac{%
\left( \partial _{u}R\right) }{R}  \notag \\
+\left( \sqrt{-g}g^{yy}\right) \frac{(\partial _{yy}S)}{S}+\left[ \partial
_{y}\left( \sqrt{-g}g^{yy}\right)\right] \frac{\left( \partial _{y}S\right)
}{S}=0,
\end{eqnarray}
to see the structure better with individual radial and angular functions. We know from the original article \cite{Wu:2011gq} that the determinant of the metric can be written as
\begin{equation}
\sqrt{-g}=(H_{1}H_{2}H_{3})^{1/3}\frac{\Sigma }{4\xi _{a}\xi_{b}(a^{2}-b^{2})},
\end{equation}
in our new coordinates and consequently we get
\begin{eqnarray}
\sqrt{-g}g^{uu} &=&\frac{X}{\xi _{a}\xi _{b}(a^{2}-b^{2})}, \\
\sqrt{-g}g^{yy} &=&\frac{Y}{\xi _{a}\xi _{b}(a^{2}-b^{2})}.
\end{eqnarray}
This simple form of the determinant which is regular at the horizon is a consequence of the metric for the horizon $g^{uu}$ that gives the pole structure. We can then rewrite the Klein-Gordon equation
\begin{equation}
-\omega ^{2}\xi _{a}\xi _{b}(a^{2}-b^{2})\sqrt{-g}g^{tt}+X\frac{(\partial
_{uu}R)}{R}+\left( \partial _{u}X\right) \frac{\left( \partial _{u}R\right)
}{R}+Y\frac{(\partial _{yy}S)}{S}+\left( \partial _{y}Y\right) \frac{\left(
\partial _{y}S\right) }{S}=0,
\end{equation}
and the problem reduces to the separation of the term without derivatives, namely
\begin{equation}
\mathcal{T}\equiv -\omega ^{2}\xi _{a}\xi _{b}(a^{2}-b^{2})\sqrt{-g}g^{tt},
\end{equation}
which is studied in the Appendix A.1 in detail. The result can be given by
\begin{equation}
\mathcal{T}=-\frac{\omega ^{2}}{4}\left( \frac{F}{X}+\frac{\xi _{a}\xi _{b}y}{%
(1-g^{2}y)}\right) ,
\end{equation}
where
\begin{equation}
F=\xi _{a}\xi _{b}f_{u}+mt_{23s}-m\chi \left( a^{2}\xi_{b}+b^{2}\xi_{a}\right) ,
\end{equation}
and the $F$ function is cubic in the radial coordinate $u$ with no dependence on the angular coordinate $y$. This fact can be seen in the expressions forming the $F$ function, given by the equations (\ref{eq:f1}-\ref{eq:f2}) of the Appendix A.1.

Therefore, the Klein-Gordon equation becomes
\begin{equation}
X\frac{(\partial _{uu}R)}{R}+\left( \partial _{u}X\right) \frac{\left(
\partial _{u}R\right) }{R}+Y\frac{(\partial _{yy}S)}{S}+\left( \partial
_{y}Y\right) \frac{\left( \partial _{y}S\right) }{S}-\frac{\omega ^{2}F}{4X}-%
\frac{\omega ^{2}\xi _{a}\xi _{b}y}{4(1-g^{2}y)}=0,
\end{equation}
and the angular and radial equations can be read as
\begin{eqnarray}
Y\frac{(\partial _{yy}S)}{S}+\left( \partial _{y}Y\right) \frac{\left(
\partial _{y}S\right) }{S}-\frac{\omega ^{2}\xi _{a}\xi _{b}y}{4(1-g^{2}y)}
&=&c_{0}, \\
X\frac{(\partial _{uu}R)}{R}+\left( \partial _{u}X\right) \frac{\left(
\partial _{u}R\right) }{R}-\frac{\omega ^{2}F}{4X} &=&-c_{0}.
\end{eqnarray}%
where$\ c_{0}$ is the separation constant.
\subsection{\it{Angular equation}}
The angular equation,
\begin{equation}
\partial _{y}\left[ Y\left( \partial _{y}S\right) \right] -\left( \frac{\omega ^{2}\xi _{a}\xi _{b}y}{4(1-g^2y)}+c_{0}\right) S=0,
\end{equation}
has four regular singularities as $\left\{ a^{2},b^{2},\frac{1}{g^{2}},\infty \right\}$ and it can be solved in terms of the general Heun's functions \cite{Ronveaux95,Birkandan:2006ac,Hortacsu:2011rr}.
\subsection{\it{Analysis of the radial equation}}
Let us deal with the radial equation now. It can be written as
\begin{equation}
\partial _{u}\left[ X\left( \partial _{u}R\right) \right] -\left( \frac{\omega ^{2}F}{4X}-c_{0}\right) R=0.
\end{equation}
and we have the horizon equation given in (\ref{eq:horeqn}). Let us then try to see if we have a structure for the radial equation as
\begin{eqnarray}
&&\partial _{u}\left[ X\left( \partial _{u}R\right) \right]   \notag \\
&&+\left( n_{1}\frac{\omega ^{2}}{4\kappa _{1}^{2}\left( u-u_{1}\right) }+n_{2}\frac{\omega ^{2}}{4\kappa _{2}^{2}\left( u-u_{2}\right) }+n_{3}\frac{\omega ^{2}}{4\kappa _{3}^{2}\left( u-u_{3}\right)}+n_{4}\omega^{2}+n_{5}\right) R=0, \label{eq:core}
\end{eqnarray}
where $\kappa_{i}$ ($i=1,2,3$) is the surface gravity associated with the horizon $u_{i}$. This equation will be the core result of this section once it is proven. It also has four regular singularities as $\{u_{1},u_{2},u_{3},\infty \}$ and it can be solved in terms of the general Heun's functions.

The theory of the Heun's equation and its confluent forms is not developed enough to yield a direct analysis of the cases involving these equations. For example, there are major issues in the connection problem of the local solutions and in the integral representations of the functions \cite{Ronveaux95}. Therefore, the lengthy outputs of the radial and angular solutions in our case would not be helpful for a further study and will not be included in this paper.

The form given in \eqref{eq:core} of the radial equation was previously found in \cite{CLI,CLII} to the cases of general rotating multi-charged black holes in maximally supersymmetric ungauged supergravities \cite{CYII,CYI} and the gauged solution \cite{Birkandan:2011fr} for three equal charges \cite{Chong05}.
\subsubsection{\it{Residua of the poles and surface gravities}}
We expect to have
\begin{eqnarray}
n_{1} &=&-g^{2}(u_{1}-u_{2})(u_{1}-u_{3}), \label{eq:n1}\\
n_{2} &=&g^{2}(u_{1}-u_{2})(u_{2}-u_{3}), \label{eq:n2}\\
n_{3} &=&-g^{2}(u_{1}-u_{3})(u_{2}-u_{3}),\label{eq:n3}
\end{eqnarray}
as in the similar case we had in a previous paper \cite{Birkandan:2011fr}. Matching the terms in the both forms, we immediately see
\begin{equation}
n_{5}=c_{0},
\end{equation}
using the zeroth power of $\omega$ in the both sides. We also get
\begin{equation}
n_{4}=-\,{\frac{\xi _{a}\xi _{b}}{4{g}^{2}},}
\end{equation}
easily using the coefficient of $u^{3}$. The remaining term to analyze is
\begin{equation}
-F+\frac{\xi _{a}\xi _{b}X}{{g}^{2}}=\frac{n_{1}X}{\kappa _{1}^{2}\left(
u-u_{1}\right) }+\frac{n_{2}X}{\kappa _{2}^{2}\left( u-u_{2}\right) }+\frac{%
n_{3}X}{\kappa _{3}^{2}\left( u-u_{3}\right) }\,.
\end{equation}
For the surface gravities we find
\begin{equation}
\kappa _{i}^{2}=\frac{n_{i}^{2}}{F_{i}},\text{ \ \ \ \ \ }(i=1,2,3),
\end{equation}
where $F_{i}\equiv F|_{u=u_{i}}$ calculated on the horizon. Then we see that the equality
\begin{equation}
-F+\frac{\xi _{a}\xi _{b}X}{{g}^{2}}=X\left( \frac{F_{1}}{n_{1}\left(
u-u_{1}\right) }+\frac{F_{2}}{n_{2}\left( u-u_{2}\right) }+\frac{F_{3}}{%
n_{3}\left( u-u_{3}\right) }\right) \,,
\end{equation}
holds after performing some algebraic manipulation.

We know from the equation (\ref{eq:kappas}) that
\begin{equation}
\kappa _{i}=\sqrt{\frac{u_{i}}{W_{i}}}\left( \frac{dX}{du}\right)|_{u=u_{i}}=n_{i}\sqrt{\frac{u_{i}}{W_{i}}}.
\end{equation}
On the other hand we know that the entropy given by the equation (\ref{eq:entropy}) on the horizon $u_{i}$ can be written as
\begin{equation}
S_i=\frac{\pi ^{2}}{2\xi _{a}\xi _{b}}\sqrt{\frac{W_{i}}{u_{i}}}=\frac{A_{i}}{4},
\end{equation}
where $A_{i}$ is the area of the black hole associated with the horizon $u_{i}$. Then we can relate the surface gravity with the horizon area
\begin{equation}
\kappa _{i}=\frac{2\pi ^{2}}{\xi _{a}\xi _{b}}\frac{n_{i}}{A_{i}},
\end{equation}
where the coefficient $\pi ^{2}$ comes from the integration over the azimuthal angles.
\section{Full Klein-Gordon equation}
We have taken zero azimuthal eigenvalues in the scalar field Ansatz to obtain the separated Klein-Gordon equation to its radial and angular parts in the previous section. In this section, we will generalize our calculation to nonzero azimuthal eigenvalues and try to get the full radial equation by showing the completion of squares of the derivative-free terms including the angular velocities associated with the azimuthal coordinates. We will also give the full form of the angular equation.

The full massless Klein-Gordon equation (\ref{eq:kg}) can be written with the Ansatz (\ref{eq:ansatz}) as
\begin{eqnarray}
&&\partial _{u}\left( \sqrt{-g}g^{uu}\partial _{u}\Phi \right) +\partial_{y}\left( \sqrt{-g}g^{yy}\partial _{y}\Phi \right)  \notag \\
&&-\omega ^{2}\sqrt{-g}g^{tt}\Phi +2\omega m_{1}\sqrt{-g}g^{t\phi}\Phi +2\omega m_{2}\sqrt{-g}g^{t\psi }\Phi  \notag \\
&&-m_{1}^{2}\sqrt{-g}g^{\phi \phi }\Phi -m_{2}^{2}\sqrt{-g}g^{\psi \psi}\Phi -2m_{1}m_{2}\sqrt{-g}g^{\phi \psi }\Phi =0.
\end{eqnarray}
The separability of the terms associated with the azimuthal coordinates is studied in the Appendix A.2. We see no contribution outside the residua of the poles in the radial part because of the linear and quadratic $u$-coordinate dependencies of the coefficients.

The derivative-free terms in the full Klein-Gordon equaton are given by
\begin{equation}
\mathcal{D}_{f} =-\sqrt{-g} \left( \omega ^{2}g^{tt}-2\omega m_{1}g^{t\phi }-2\omega m_{2}g^{t\psi }+m_{1}^{2}g^{\phi \phi }+m_{2}^{2}g^{\psi \psi}+2m_{1}m_{2}g^{\phi \psi }\right). \label{eq:dfree}
\end{equation}
On the other hand, we believe that we have a structure like
\begin{equation}
\overset{\#poles}{\underset{i=1}{\sum }}n_{i}\frac{\alpha _{i}^{2}}{u-u_{i}},
\end{equation}
for the derivative-free terms associated with the poles, based on the results obtained in the articles \cite{CLI,CLII,Birkandan:2014cva}, where
\begin{eqnarray}
n_i&=&\frac{dX}{du}|_{u=u_{i}},\\
\alpha _{i}&=&\frac{\omega -m_{1}\Omega _{\phi i}-m_{2}\Omega _{\psi i}}{2\kappa _{i}}. \label{eq:alphak}
\end{eqnarray}
\subsection{\it{Residua of the poles and angular velocities}}
Let us start with the general form of a stationary metric with bi-azimuthal symmetry, namely
\begin{equation}
ds^{2}=g_{tt}dt^{2}+g_{uu}du^{2}+g_{\theta \theta }d\theta ^{2}+g_{\phi \phi}d\phi ^{2}+g_{\psi \psi }d\psi ^{2}+2g_{t\phi }dtd\phi +2g_{t\psi }dtd\psi+2g_{\phi \psi }d\phi d\psi,
\end{equation}
with three Killing vectors $\partial_{t}, \partial_{\phi}$ and $\partial_{\psi}$ \cite{Myers:1986un,Aliev:2004ec}.

The definitions of angular velocities for such a stationary solution is
\begin{eqnarray}
\Omega _{\phi i} &\equiv &\frac{g^{t\phi }}{g^{tt}}|_{u=u_{i}}, \\
\Omega _{\psi i} &\equiv &\frac{g^{t\psi }}{g^{tt}}|_{u=u_{i}},
\end{eqnarray}
evaluated at the horizon $u=u_{i}$. The angular velocities are $\theta$ independent on the horizon since the Killing vector $\partial_t+\Omega_\phi\partial_\phi+\Omega_\psi\partial_\psi$ becomes null. We can choose $\theta=0$ to simplify our calculations. Then we have
\begin{eqnarray}
\Omega _{\phi i} &=&\left( -\frac{g_{t\phi }}{g_{\phi \phi }}+\frac{g_{t\psi}g_{\phi \psi }}{g_{\psi \psi }g_{\phi \phi }}\right) |_{u=u_{i},\theta =0}, \\
\Omega _{\psi i} &=&\left( -\frac{g_{t\psi }}{g_{\psi \psi }}\right)|_{u=u_{i},\theta =0}.
\end{eqnarray}
It is practical to keep the details of the calculations in the Appendix B in which we used special values for the $\theta$ coordinate without losing generality of the results. The most important point we used in these calculations is that the determinant being regular at the horizon. This fact enforces some constraints on the metric elements that simplifies the expressions yielding the angular velocities on the horizon. At $\theta=0$ we find
\begin{equation}
g_{tt}g_{\psi \psi }-g_{t\psi }^{2}=0,
\end{equation}
using the determinant of the metric. Accordingly we have
\begin{eqnarray}
\frac{g^{\phi \psi }}{g^{tt}}|_{u=u_{i},\theta =0} &=&-\frac{g_{tt}g_{\phi \psi}-g_{t\phi }g_{t\psi }}{g_{\phi \phi }g_{\psi \psi }}|_{u=u_{i},\theta =0} \\
&=&\Omega _{\phi i}\Omega _{\psi i},
\end{eqnarray}
and similarly we obtain
\begin{eqnarray}
\Omega _{\phi i}^{2} &=&\frac{g^{\phi \phi }}{g^{tt}}|_{u=u_{i}}, \\
\Omega _{\psi i}^{2} &=&\frac{g^{\psi \psi }}{g^{tt}}|_{u=u_{i}}.
\end{eqnarray}
Therefore, we find a general correspondence between (\ref{eq:dfree}) and (\ref{eq:alphak}) expressions which is needed to have the structure we proposed using these results. The surface gravity in the equation (\ref{eq:alphak}) comes from the $\sqrt{-g}g^{tt}$ part as we have seen in the previous section.

Finally, we obtain the full radial equation as
\begin{eqnarray}
&&\partial _{u}\left[ X\left( \partial _{u}R\right) \right]   \notag \\
&&+\bigg( n_{1}\frac{\left( \omega -m_{1}\Omega _{\phi 1}-m_{2}\Omega_{\psi 1}\right) ^{2}}{4\kappa _{1}^{2}\left( u-u_{1}\right) } +n_{2}\frac{\left( \omega -m_{1}\Omega_{\phi 2}-m_{2}\Omega_{\psi 2}\right)^{2}}{4\kappa _{2}^{2}\left( u-u_{2}\right) } \notag \\
&&+n_{3}\frac{\left( \omega -m_{1}\Omega _{\phi 3}-m_{2}\Omega_{\psi 3}\right) ^{2}}{4\kappa _{3}^{2}\left( u-u_{3}\right) }+n_{4}\omega^{2}+n_{5} \bigg) R=0, \label{eq:fullrad}
\end{eqnarray}
and the angular equation,
\begin{equation}
\partial _{y}\left[Y\left( \partial _{y}S\right) \right] -\left( \frac{\omega ^{2}\xi _{a}\xi _{b}y}{4(1-g^2y)}+\frac{{m_1^2 \xi_a (a^2-b^2)}}{4(a^2-y)}+\frac{{m_2^2 \xi_b (a^2-b^2)}}{4(y-b^2)}+c_{0}\right) S=0, \label{eq:fullang}
\end{equation}
where the surface gravities $\kappa_i$ given in (\ref{eq:kappas}), the angular velocities $\Omega_{\phi i}$ and $\Omega_{\psi i}$ in (\ref{eq:Omegaa},\ref{eq:Omegab}), $n_i$ constants in (\ref{eq:n1}-\ref{eq:n3}) and $c_0$ is the separation constant. The equations (\ref{eq:fullrad}) and (\ref{eq:fullang}) are the core results of the paper.

The singularity structures of the both equations remain the same and they can still be solved in terms of general Heun's functions.
\section{Conclusions}
In this paper, we discussed the separability of the massless Klein-Gordon equation and probed the singularities for a general stationary, axisymmetric metric with orthogonal transitivity, and applied our method to the Wu black hole,  the most general  general nonextremal rotating charged AdS black hole solution in five-dimensional $U(1)^{3}$ gauged supergravity  theory.
\begin{itemize}
\item The separation problem of the Klein-Gordon equation is reduced to the separation of the derivative-free terms of the equation with the problematic piece $\Sigma=u+y$ which involves both radial coordinate $u$ and polar angular  coordinate $y$. We carefully studied the coefficients of this piece and  succeeded in obtaining separated terms for the whole inverse metric elements.  This resulted in the emergence of a Killing-St\"ackel tensor which guarantees the separability of the Klein-Gordon equation. The main point we used throughout our calculations is the fact that the determinant of the metric has a simple form and it is regular at the horizon.
\item Probing each horizon singularity, we showed that the residua of the singularities for each horizon are associated in general with a specific combination of the surface gravity and the angular velocities at the associated horizon. We showed that this is true for all stationary, axisymmetric metric with orthogonal transitivity.
\item We firstly used zero eigenvalues for the azimuthal coordinates $\phi$ and $\psi$ in the scalar field Ansatz to show the separability. Then turning the azimuthal eigenvalues on, we generalized the solution to the full separable Klein-Gordon equation which includes the angular velocities as well.
\item Both the radial and  the angular equations  can be solved in terms of the general Heun's functions with their four regular singularities.
\end{itemize}
This study is a starting point for further investigations such as calculation of Green's functions, near-extremal cases, small cosmological constant limit and limits related to subtracted geometry.
\acknowledgments
We would like to thank Finn Larsen and Gary Gibbons for discussions. TB would like to thank the Department of Physics and Astronomy at the University of Pennsylvania for hospitality during the early course of this work. The research of TB is supported by TUBITAK, the Scientific and Technological Council of Turkey, Istanbul Technical University [\.{I}T\"{U}\ BAP 37519] and High Energy Theory Fund of the Department of Physics and Astronomy at the University of Pennsylvania. MC would like to thank the Department of Physics at the Istanbul Technical University for hospitality during the course of this work. MC's visit to Istanbul Technical University was supported by TUBITAK ``2221-Fellowship for Visiting Scientists" with number 1059B211400897. MC is also supported by the DOE Grant DOE-EY-76-02-3071, the Fay R. and Eugene L. Langberg Endowed Chair, the Slovenian Research Agency (ARRS), and the Simons Foundation Fellowship No. 267489. MC would also like to thank KITPC Beijing,  CERN Theory Division and MITP Mainz for hospitality during the later stages of the work.
\appendix
\section{Separation of the inverse metric elements}
The problem of separation of the Klein-Gordon equation reduces to the separation of the inverse metric elements. For the zero azimuthal eigenvalues we only have the term with $g^{tt}$ and for the full equation we need to show the whole terms containing the azimuthal angles.
\subsection{\it{Separation of the derivative-free term for the zero azimuthal eigenvalues}}
The problem of separation of the Klein-Gordon equation reduces to the separation of the term
\begin{equation}
\mathcal{T}\equiv -\omega ^{2}\xi _{a}\xi _{b}(a^{2}-b^{2})\sqrt{-g}g^{tt},
\end{equation}
and using the the metric determinant we have
\begin{equation}
\mathcal{T}\equiv -\frac{4\omega ^{2}\xi _{a}^{2}\xi _{b}^{2}(a^{2}-b^{2})^{2}}{%
(H_{1}H_{2}H_{3})^{1/3}\Sigma }g_{uu}g_{yy}\left( g_{\phi \phi }g_{\psi \psi
}-g_{\phi \psi }^{2}\right) .
\end{equation}
Using
\begin{equation}
H_{i}=1+\frac{2ms_{i}^{2}}{\Sigma }=\frac{\Sigma +2ms_{i}^{2}}{\Sigma }\equiv \frac{h_{i}}{\Sigma },
\end{equation}
and remembering $\sin ^{2}\theta =\frac{a^{2}-y}{a^{2}-b^{2}},\cos ^{2}\theta =\frac{y-b^{2}}{a^{2}-b^{2}}$, with the definitions
\begin{eqnarray}
k_{i\phi } &\equiv &\frac{\xi _{a}}{\sin ^{2}\theta }K_{i\phi }, \\
k_{i\psi } &\equiv &\frac{\xi _{b}}{\cos ^{2}\theta }K_{i\psi },
\end{eqnarray}
we get
\begin{equation}
\mathcal{T}=-\frac{\omega ^{2}}{4X(1-g^{2}y)\Sigma ^{2}}\left(
T_{1}+T_{2}+T_{3}+T_{4}\right) ,
\end{equation}
where
\begin{eqnarray}
T_{1} &=&(u+a^{2})(u+b^{2})\xi _{a}\xi _{b}(h_{1}h_{2}h_{3}), \\
T_{2} &=&(u+a^{2})2m\xi _{a}\cos ^{2}\theta \left( \eta _{1}k_{1\psi
}^{2}h_{2}h_{3}+\eta _{2}k_{2\psi }^{2}h_{1}h_{3}+\eta _{3}k_{3\psi
}^{2}h_{1}h_{2}\right) , \\
T_{3} &=&(u+b^{2})2m\xi _{b}\sin ^{2}\theta \left( \eta _{1}k_{1\phi
}^{2}h_{2}h_{3}+\eta _{2}k_{2\phi }^{2}h_{1}h_{3}+\eta _{3}k_{3\phi
}^{2}h_{1}h_{2}\right) , \\
T_{4} &=&4m^{2}(h_{1}h_{2}h_{3})\sin ^{2}\theta \cos ^{2}\theta \left[
\left( \overset{3}{\underset{i=1}{\sum }}\frac{\eta _{i}k_{i\phi }^{2}}{h_{i}%
}\right) \left( \overset{3}{\underset{i=1}{\sum }}\frac{\eta _{i}k_{i\psi
}^{2}}{h_{i}}\right) \right.  \notag \\
&&\left. -\left( \overset{3}{\underset{i=1}{\sum }}\frac{\eta _{i}k_{i\phi
}k_{i\psi }}{h_{i}}\right) ^{2}\right] ,
\end{eqnarray}
and using $u=\Sigma -y$ and
\begin{eqnarray}
u+a^{2} &=&\Sigma +\left( a^{2}-b^{2}\right) \sin ^{2}\theta , \\
u+b^{2} &=&\Sigma -\left( a^{2}-b^{2}\right) \cos ^{2}\theta ,
\end{eqnarray}
we observe
\begin{eqnarray}
T_{1} &=&T_{10}+T_{11}\Sigma +T_{12}\Sigma ^{2}+T_{13}\Sigma
^{3}+T_{14}\Sigma ^{4}+T_{15}\Sigma ^{5}, \\
T_{2} &=&T_{20}+T_{21}\Sigma +T_{22}\Sigma ^{2}+T_{23}\Sigma ^{3}, \\
T_{3} &=&T_{30}+T_{31}\Sigma +T_{32}\Sigma ^{2}+T_{33}\Sigma ^{3}, \\
T_{4} &=&T_{40}+T_{41}\Sigma .
\end{eqnarray}
We need to have zero as the coefficients of $\Sigma ^{0}$ and $\Sigma ^{1}$ powers as we have $\Sigma ^{2}$ in the common denominator. The Maple computation including all metric parameters shows that these coefficients are indeed zero and now we have
\begin{eqnarray}
t_{1} &=&T_{12}+T_{13}\Sigma +T_{14}\Sigma ^{2}+T_{15}\Sigma
^{3}=t_{1u}+t_{1y}y, \\
t_{2} &=&T_{22}+T_{23}\Sigma , \\
t_{3} &=&T_{32}+T_{33}\Sigma ,
\end{eqnarray}
and consequently
\begin{equation}
\mathcal{T}=-\frac{\omega ^{2}}{4X(1-g^{2}y)}\left( t_{1}+t_{2}+t_{3}\right) .
\end{equation}
After some algebraic manipulation we get
\begin{equation}
\mathcal{T}=-\frac{\omega ^{2}}{4}\left( \frac{F}{X}+\frac{\xi _{a}\xi _{b}y}{(1-g^{2}y)}\right) ,
\end{equation}
where
\begin{eqnarray}
F &=&\xi _{a}\xi _{b}f_{u}+mt_{23s}-m\chi \left( a^{2}\xi _{b}+b^{2}\xi_{a}\right) , \\
G &=&g^{2}f_{u}+2m\chi +g_{u}=X,
\end{eqnarray}
and
\begin{eqnarray}
f_{u} &=&\frac{t_{1u}}{\xi _{a}\xi _{b}}, \label{eq:f1} \\
g_{u} &=&\frac{t_{1y}}{\xi _{a}\xi _{b}}, \\
t_{23d} &=&\frac{1}{2m}\left( \frac{t_{2}}{\xi _{a}\cos ^{2}\theta }-\frac{%
t_{3}}{\xi _{b}\sin ^{2}\theta }\right) , \\
t_{23s} &=&\frac{1}{2m}\left( \frac{t_{2}}{\xi _{a}\cos ^{2}\theta }+\frac{%
t_{3}}{\xi _{b}\sin ^{2}\theta }\right) , \\
\chi &=&\frac{t_{23d}}{a^{2}-b^{2}},
\end{eqnarray}
and the explicit forms of the expressions included in $F$ can be given as
\begin{eqnarray}
f_{u} &=&8\,{m}^{3}s_{1}^{2}s_{2}^{2}s_{3}^{2}  \nonumber \\
&&+4m^{2}[(\,{a}^{2}+b^{2})\left(
s_{1}^{2}s_{2}^{2}+s_{1}^{2}s_{3}^{2}+\,s_{2}^{2}s_{3}^{2}\right) ]
\nonumber \\
&&+2\,{a}^{2}{b}^{2}m(s_{1}^{2}+s_{2}^{2}+s_{3}^{2})  \nonumber \\
&&+2m[2m(s_{1}^{2}s_{2}^{2}+s_{1}^{2}s_{3}^{2}+s_{2}^{2}s_{3}^{2})+(\,{a}%
^{2}+b^{2})(s_{1}^{2}+s_{2}^{2}+s_{3}^{2})+\frac{{a}^{2}{b}^{2}}{2m}]u  \nonumber \\
&&+[2\,m\left( s_{1}^{2}+s_{2}^{2}+s_{3}^{2}\right) +{a}^{2}+{b}^{2}]{u}^{2}+%
{u}^{3}, \\
t_{23d} &=&-2m[\,\left( s_{2}^{2}+s_{3}^{2}\right) \left( k_{1\phi
}^{2}-k_{1\psi }^{2}\right) \eta _{1}+\left( s_{1}^{2}+s_{3}^{2}\right)
\left( k_{2\phi }^{2}-k_{2\psi }^{2}\right) \eta _{2}  \nonumber \\
&&+\left( s_{1}^{2}+s_{2}^{2}\right) \left( k_{3\phi }^{2}-k_{3\psi
}^{2}\right) \eta _{3}]+\left[ {a}^{2}k_{1\psi }^{2}-{b}^{2}k_{1\phi
}^{2}+\left( k_{1\psi }^{2}-k_{1\phi }^{2}\right) u\right] \eta _{1} \\
&&+\left[ {a}^{2}k_{2\psi }^{2}-{b}^{2}k_{2\phi }^{2}+\left( k_{2\psi
}^{2}-k_{2\phi }^{2}\right) u\right] \eta _{2}+\,\left[ {a}^{2}k_{3\psi
}^{2}-{b}^{2}k_{3\phi }^{2}+\left( k_{3\psi }^{2}-k_{3\phi }^{2}\right) u%
\right] \eta _{3},  \nonumber \\
t_{23s} &=&2m[\,\left( s_{2}^{2}+s_{3}^{2}\right) \left( k_{1\phi
}^{2}+k_{1\psi }^{2}\right) \eta _{1}+\,\left( s_{1}^{2}+s_{3}^{2}\right)
\left( k_{2\phi }^{2}+k_{2\psi }^{2}\right) \eta _{2}  \nonumber \\
&&+\,\left( s_{1}^{2}+s_{2}^{2}\right) \left( k_{3\phi }^{2}+k_{3\psi
}^{2}\right) \eta _{3}]+\left[ {a}^{2}k_{1\psi }^{2}+{b}^{2}k_{1\phi
}^{2}+\left( k_{1\phi }^{2}+k_{1\psi }^{2}\right) u\right] \eta _{1} \label{eq:f2} \\
&&+\left[ {a}^{2}k_{2\psi }^{2}+{b}^{2}k_{2\phi }^{2}+\left( k_{2\phi
}^{2}+k_{2\psi }^{2}\right) u\right] \eta _{2}+\left[ \,{a}^{2}k_{3\psi
}^{2}+{b}^{2}k_{3\phi }^{2}+\left( k_{3\phi }^{2}+k_{3\psi }^{2}\right) u%
\right] \eta _{3},  \nonumber
\end{eqnarray}
to see the structure more clearly.
\subsection{\it{Separation of the azimuthal inverse metric elements}}
We need to see the separability of the whole elements in the full Klein-Gordon equation involving the inverse metric components associated with the azimuthal angles.

Repeating the same method we applied in the previous section we obtain that the mixed terms show the same linear behavior in $u$ as
\begin{equation}
\sqrt{-g}g^{\tilde{\mu} \tilde{\nu}}=\frac{C_{1 \tilde{\mu} \tilde{\nu}}u+C_{2 \tilde{\mu} \tilde{\nu}}}{X}
\end{equation}
where $g^{\tilde{\mu} \tilde{\nu}}$ can be one of the \{$g^{t\phi}$, $g^{t\psi}$, $g^{\phi \psi}$\} and $C_{1 \tilde{\mu} \tilde{\nu}}$ and $C_{2 \tilde{\mu} \tilde{\nu}}$ have no dependence on $u$ or $y$ coordinates. Thus, these terms do not contribute to the angular equation or to the asymptotic term of the radial equation.

We have a different behavior for the $g^{\phi \phi }$ and $g^{\psi \psi }$ components as
\begin{eqnarray}
\sqrt{-g}g^{\phi \phi } &= &-\frac{\cos ^{2}\theta \left( p_{21}\cos^{2}\theta +p_{01}\right) }{4X\xi_a \xi_b^3 (a^2-b^2)^2\sin ^{2}\theta \cos ^{2}\theta } \\
&= &\frac{p_{21}}{4X\xi_a \xi_b^3 (a^2-b^2)^2}-\frac{1}{4 (a^2-b^2) \xi_b \sin ^{2}\theta },
\end{eqnarray}
where $p_{01}$ and $p_{21}$ are expressions with no dependence on the $y$ coordinate and
\begin{equation}
p_{21}= C_{31}u^{2}+C_{41}u+C_{51},
\end{equation}
again with no $u$ or $y$ dependence for the $C_{ij}$ expressions. This separation is realized as $(p_{01}+p_{21})=\xi_a \xi_b^2 (a^2-b^2)X$. The same analysis holds for $g^{\psi \psi }$ with $a\rightarrow b$ and $\cos \theta \rightarrow \sin \theta $ and
\begin{equation}
\sqrt{-g}g^{\psi \psi }= \frac{p_{22}}{4X\xi_a^3 \xi_b (a^2-b^2)^2}-\frac{1}{4 (a^2-b^2) \xi_a \cos ^{2}\theta },
\end{equation}
with the same quadratic dependence structure for $p_{22}$ as in $p_{21}$.

If one has all separable $\sqrt{-g}g^{\mu \nu }$ components for a stationary, axisymmetric metric with orthogonal transitivity which agrees with our case, it is guaranteed that the Klein-Gordon equation is separable with the existence of a conformal Killing-St\"ackel tensor \cite{Keeler:2012mq}.
\section{Angular velocities on the horizon}
The angular velocities are $\theta$ independent on the horizon. We can choose $\theta=0$ in order to simplify our calculations but the results could be regarded as general. Then we have
\begin{eqnarray}
\Omega _{\phi i} &=&\left( -\frac{g_{t\phi }}{g_{\phi \phi }}+\frac{g_{t\psi}g_{\phi \psi }}{g_{\psi \psi }g_{\phi \phi }}\right) |_{u=u_{i},\theta =0}, \\
\Omega _{\psi i} &=&\left( -\frac{g_{t\psi }}{g_{\psi \psi }}\right)|_{u=u_{i},\theta =0}.
\end{eqnarray}
On the horizon we have $g_{uu}\rightarrow \infty$ with a regular value of the determinant. Consequently, the expressions in the determinant except this term should be zero. One needs to use the leading orders in the expansion around $\theta =0$ and see the metric components in our case are found to behave like
\begin{eqnarray}
g_{\phi \phi } &=&g_{0\phi \phi }\sin ^{2}\theta +g_{1\phi \phi }\sin^{4}\theta , \\
g_{t\phi } &=&g_{0t\phi }\sin ^{2}\theta , \\
g_{tt} &=&g_{0tt}\sin ^{2}\theta +g_{2tt}, \\
g_{\psi \psi } &=&g_{0\psi \psi }\cos ^{2}\theta +g_{1\psi \psi }\cos^{4}\theta , \\
g_{t\psi } &=&g_{0t\psi }\cos ^{2}\theta , \\
g_{\phi \psi } &=&g_{0\phi \psi }\sin ^{2}\theta \cos ^{2}\theta ,
\end{eqnarray}
where the coefficients $g_{0\phi \phi }$, etc. are regular at $\theta =0$. Around $\theta =0$, one sees that the determinant becomes
\begin{equation}
g=g_{uu}g_{\theta \theta }g_{0\phi \phi }\sin ^{2}\theta \left[g_{2tt}(g_{0\psi \psi }+g_{1\psi \psi })-g_{0t\psi }^{2}\right] +\mathcal{O}(\sin ^{4}\theta )
\end{equation}
and thus
\begin{equation}
g_{2tt}(g_{0\psi \psi }+g_{1\psi \psi })-g_{0t\psi }^{2}=0,
\end{equation}
or
\begin{equation}
g_{tt}g_{\psi \psi }-g_{t\psi }^{2}=0, \label{eq:fromdet}
\end{equation}
at $\theta=0$. We also have
\begin{eqnarray}
\frac{g^{\phi \psi }}{g^{tt}}|_{u=u_{i},\theta =0} &=&-\frac{g_{tt}g_{\phi \psi}-g_{t\phi }g_{t\psi }}{g_{\phi \phi }g_{\psi \psi }}|_{u=u_{i},\theta =0} \\
&=&\left( -\frac{g_{t\psi }}{g_{\psi \psi }}\right) |_{u=u_{i},\theta =0}\left( -\frac{g_{t\phi }}{g_{\phi \phi }}+\frac{g_{t\psi }}{g_{\psi \psi }}\frac{g_{\phi \psi }}{g_{\phi \phi }}\right) |_{u=u_{i},\theta=0,}
\end{eqnarray}
using equation (\ref{eq:fromdet}). The expression in the first bracket is $\Omega _{\psi i}$ and the second bracket gives $\Omega _{\phi i}$. We should remember that using $\theta =0$ is only for the simplification of our derivations and the expressions are valid for a general $\theta$ value. Therefore we can write
\begin{equation}
\Omega _{\phi i}\Omega _{\psi i}=\frac{g^{\phi \psi }}{g^{tt}}|_{u=u_{i}}.
\end{equation}
Similarly one can show that
\begin{eqnarray}
\Omega _{\phi i} &=&\frac{g^{t\phi }}{g^{tt}}|_{u=u_{i}}, \label{eq:omegaak}\\
\Omega _{\psi i} &=&\frac{g^{t\psi }}{g^{tt}}|_{u=u_{i}}, \label{eq:omegabk}\\
\Omega _{\phi i}^{2} &=&\frac{g^{\phi \phi }}{g^{tt}}|_{u=u_{i}}, \label{eq:omegaaksq}\\
\Omega _{\psi i}^{2} &=&\frac{g^{\psi \psi }}{g^{tt}}|_{u=u_{i}}, \label{eq:omegabksq}
\end{eqnarray}
are satisfied. One should note that we need to use $\theta =\frac{\pi }{2}$ from the beginning in our calculations to show the equivalence (\ref{eq:omegaaksq}). Different values of $\theta $ supplies different expressions for the part of the determinant that should be zero on the horizon and these expressions simplifies our calculations.

\end{document}